\documentclass[12pt,twoside]{article}
\usepackage[mathscr]{eucal}
\usepackage{amsmath,amsfonts,amssymb,amsthm,mathabx,empheq}
\usepackage{times}
\usepackage{pdfsync}
\usepackage{cite}
\usepackage{url}
\usepackage{hyperref}
\usepackage{tensor}
\usepackage{color}

\advance\textheight\topskip
\allowdisplaybreaks[1]

\newcommand\td{\text{d}}
\newcommand\cO{{\cal O}}
\newcommand{\p}{\partial}

\newcommand{\be}{\begin{equation}}
\newcommand{\ee}{\end{equation}}
\newcommand{\bea}{\begin{eqnarray}}
\newcommand{\eea}{\end{eqnarray}}

\def\bz{\bar z}
\def\ga{\gamma_{z\bz}}
\def\sga{\sqrt\gamma_{z\bz}}
\def\pz{\p_{z}}
\def\pzb{\p_{\bz}}

\newcommand\Preind[3]{\vphantom{#3}#1#2#3}
\newcommand*\xbar[1]{%
  \hbox{%
    \vbox{%
      \hrule height 0.5pt 
      \kern0.3ex
      \hbox{%
        \kern-0.0em
        \ensuremath{#1}%
        \kern-0.0em
      }%
    }%
  }%
}

\DeclareFontFamily{OT1}{rsfs}{} \DeclareFontShape{OT1}{rsfs}{m}{n}{
<-7> rsfs5 <7-10> rsfs7 <10-> rsfs10}{}
\DeclareMathAlphabet{\mycal}{OT1}{rsfs}{m}{n}

\begin{document}
\title{Remarks on infinite towers of gravitational memories}

\author{Pujian Mao}

\date{}

\def\mytitle{Remarks on infinite towers of gravitational memories}

\pagestyle{myheadings} \markboth{\textsc{\small P.~Mao}} {\textsc{\small infinite towers of gravitational memories}}
\addtolength{\headsep}{4pt}

\begin{centering}

  \vspace{1cm}

  \textbf{\Large{\mytitle}}

  \vspace{1.5cm}

  {\large Pujian Mao}

\vspace{.5cm}

\vspace{.5cm}
\begin{minipage}{.9\textwidth}\small \it  \begin{center}
     Center for Joint Quantum Studies and Department of Physics,\\
     School of Science, Tianjin University, 135 Yaguan Road, Tianjin 300350, China
 \end{center}
\end{minipage}

\end{centering}

\vspace{1cm}

\begin{center}
\begin{minipage}{.9\textwidth}
  \textsc{Abstract}. An infinite tower of gravitational memories was proposed in \cite{Compere:2019odm} by considering the matter-induced vacuum transition in the impulsive limit. We give an alternative realization of the infinite towers of gravitational memories in Newman-Penrose formalism. We also demonstrate that the memories at each order can be associated to the same supertranslation instead of infinite towers of supertranslations or superrotations.
 \end{minipage}
\end{center}
\thispagestyle{empty}


\section{Introduction}

Gravitational memory \cite{memory,Braginsky:1986ia,1987Natur,Christodoulou:1991cr} obtained rewed attentions in recent years since its intrinsic connections to BMS supertranslations and soft graviton theorem were revealed \cite{Strominger:2014pwa}. The gravitational memory formula is a Fourier transformation of soft graviton theorem. While the memory effect is a consequence of the fact that the gravitational radiation induces transitions among two different vacua that are connected by BMS supertranslations. The triangle relation \cite{Strominger:2017zoo} has accumulated considerable evidence in its favor in particular in discovering new features of gravitational memory. In  \cite{Compere:2019odm}, it was reported that infinite towers of gravitational memory effects can be generated by matter-induced transitions (see also \cite{Hamada:2018vrw} from the soft theorem side).

In this paper, we show that the infinite towers of gravitational memories can be derived in the Newman-Penrose (NP) formalism \cite{Newman:1961qr}. The gravitational wave is characterized by the asymptotic shear $\sigma^0$ and $\xbar\sigma^0$ in the NP formalism. The infinite towers of gravitational memories are derived from the evolution of the components of the Weyl tensor. The memory effect at each order is encoded in different choices of $\sigma^0$ and $\xbar\sigma^0$. The infinite towers of gravitational memories can be associated to the unique BMS supertranslation rather than infinite towers of supertranslations and superrotations proposed in \cite{Compere:2019odm}. There are infinite conserved quantities in NP formalism at each order \cite{Newman:1968uj} which can be associated to the unique supertranslation charge \cite{Conde:2016rom}. The gravitational memories are equivalent to soft graviton theorems at the first two orders \cite{Strominger:2014pwa,Pasterski:2015tva}. We show that the derivation of equivalence for the first two orders is not valid at the third order. Thus the connection beyond the second order is still obscure. In the present work, we are restricted in linearized gravitational theory. However the nonlinearities can be captured from the contribution of the gravitational wave burst's gravitons \cite{Thorne:1992sdb}.

\section{Infinite towers of gravitational memories in Newman-Penrose formalism}

The gravitational memory effect is a relative displacement of nearby observers. The displacement of nearby observers can be derived from the geodesic deviation \cite{Flanagan:2005yc}. Consider a geodesic $x^\mu = z^\mu(\tau)$, where $\tau$ is the proper time, with four velocity $u^\mu(\tau)$. Suppose there is a nearby geodesic
$x^\mu(\tau) =  z^\mu(\tau) + L^\mu(\tau)$, where $ L^\mu(\tau)$ is small and purely spatial $L^\mu u_\mu = 0$. The coordinate displacement $ L^\mu(\tau)$ can be considered as a
vector on the first geodesic to first order in $ L^\mu(\tau)$. Space-time curvature causes the
separation vector $ L^\mu(\tau)$ to change with time with an acceleration determined by the geodesic deviation equation
\be
u^\mu \nabla_\mu (u^\nu \nabla_\nu L^\alpha)=-{R^\alpha}_{\mu\nu\beta}u^\mu  L^\nu u^\beta.
\ee
The infinite towers of memory observable in \cite{Compere:2019odm} is a displacement memory effect that derived from the geodesic deviation equation at each order in the $\frac1r$ expansion. In the NP formalism, $\sigma$ and $\xbar\sigma$ are the complex shear of the null geodesic generator $V=\frac{\p}{\p r}$ and measure its geodesic deviation. In the linearized theory, they are controlled by the radial NP equation\footnote{Here and in the following, we only list half of the equations while the complex conjugate of those equations should be noticed.}
\be
\p_r \sigma = -2 \frac{\sigma}{r} + \Psi_0\,.
\ee
Hence
\be\label{geodesic deviation}
\sigma=\frac{\sigma^0}{r^2} - \sum_{n=0}^{\infty} \frac{1}{n+2} \frac{\Psi_0^n}{r^{n+4}}\,,
\ee
where $\Psi_0$ is given as initial data\footnote{We have assumed the analyticity of the Newman-Penrose solution space with respect to the
$r$ coordinate. However, consistent solution space does exist by relaxing the analyticity of $r$, for instance, the polyhomogeneous solution space including logarithmic terms in $r$ \cite{ValienteKroon:1998vn}. If there are memories associated with the logarithmic terms is still an open question that needs to be addressed in the future.}
\be
\Psi_0 = \sum_{n=0}^\infty \frac{\Psi_0^n}{r^{n+5}}\,.
\ee
Considering the empty-space case, the evolution of the components of the Weyl tensor at each order are \cite{Newman:1968uj}
\begin{align}
&\p_u\Psi_2^0=-\eth^2\p_u\xbar\sigma^0\,,\label{p20}\\
&\p_u\Psi_1^0=-\eth\Psi_2^0\,,\label{p10}\\
&\p_u\Psi_0^0=-\eth\Psi_1^0\,,\label{p00}\\
&(n+1)\p_u\Psi_0^{n+1}=-\left[\xbar\eth\eth+(n+5)n\right]\Psi_0^n\;\;\;\;(n\geq0)\,,\label{p0j}
\end{align}
where $\Psi^0_i\,(i=1,2)$ are at order $\cO(r^{i-5})$. We use $(u,r,z,\bz)$ coordinates where $z=e^{i\phi}\cot\frac{\theta}{2},\,\bz=e^{-i\phi}\cot\frac{\theta}{2}$ are the standard stereographic coordinates. The operators $\eth$ and $\xbar\eth$ are defiend in Appendix \ref{eth}.

In the NP formalism, the gravitational wave is characterized by $\sigma^0$ and $\xbar\sigma^0$. The time evolution of the components of the Weyl tensor in \eqref{p20}-\eqref{p0j} are completely determined by $\xbar\sigma^0$ and its time integrations
\begin{align}
&\Psi_2^0\sim-\eth^2 \xbar\sigma^0\,,\label{leading}\\
&\Psi_1^0\sim\eth^3\int\,\td u' \xbar\sigma^0\,,\label{subleading}\\
&\Psi_0^0\sim-\eth^4\int\,\td u' \td u'' \xbar\sigma^0\,,\label{subsubleading}\\
&\Psi_0^{n+1}\sim(-)^{n}\prod_{k=0}^n\left[\frac{1}{k+1}[\xbar\eth\eth+(k+5)k]\int \td u_{k+1}\right] \eth^4 \int \td u' \td u'' \, \xbar \sigma^0 \,.\label{subs}
\end{align}
The memories at different orders are completely determined by the same $\xbar\sigma^0$. The independence of each memory is encoded in different choice of $\xbar\sigma^0$. For instance, $\xbar\sigma^0=\Theta(u)f(z,\bz)$ will induce a permanent change in $\xbar\sigma^0$ which will lead to a leading memory.\footnote{One should not worry about the divergent in the u-integration of $\xbar\sigma^0$. This can be resolved by choosing $f(z,\bz)$ with some anti-holomorphic properties respect to $\eth$ operator, i.e. $\eth^3 f(z,\bz)=0$.} Then $\xbar\sigma^0=\delta(u)f(z,\bz)$ will lead to a subleading memory where a permanent change happens in $\int\td u'\,\xbar\sigma^0$ instead of $\xbar\sigma^0$. Similarly, the nth order memory can be associated to $\xbar\sigma^0=\frac{\td \Theta(u)}{\td^{n-1} u} f(z,\bz)$. The memories derived from \eqref{subsubleading} and \eqref{subs} are nothing but the higher order terms in \eqref{geodesic deviation}. However the memory derived from \eqref{subleading} is somewhat different since there is a gap in the expansion in \eqref{geodesic deviation} and $\Psi_1^0$ is not in the expansion. Nevertheless the observational effect of the subleading memory is a time delay \cite{Pasterski:2015tva}.

\section{The relation to supertranslations}
Memories are associated to asymptotic symmetries. The infinite towers of gravitational memories in \cite{Compere:2019odm} are associated to infinite towers of supertranslations and superrotations. However there is only one unique supertranslation \cite{Barnich:2011ty,Conde:2016rom} in the NP formalism in the broadly used Newman-Unti gauge \cite{Newman:1962cia}. Nonetheless the infinite towers of gravitational memories defined in \eqref{leading}-\eqref{subs} can be associated to the unique supertranslation. The action of the supertranslation on $\xbar\sigma^0$ is
\be
\delta_{st}\xbar\sigma^0=-\xbar\eth^2 T(\theta,\phi)\,.
\ee
where $T(\theta,\phi)$ defines the supertranslation. The leading memory is encoded in the transition
\be
\xbar\sigma^0 + \Theta(u)\delta_{st}\xbar\sigma^0.
\ee
The subleading transition is
\be
\xbar\sigma^0+\delta(u)\delta_{st}\xbar\sigma^0.
\ee
This transition can be understood as two supertranslations in the following form
\be
\xbar\sigma^0 + \lim_{\delta u \rightarrow 0}\left(\frac{\Theta(u)}{\delta u}-\frac{\Theta(u+\delta u)}{\delta u}\right)\delta_{st}\xbar\sigma^0.
\ee
Then the nth order transition is
\be
\xbar\sigma^0 + \frac{\td \Theta(u)}{\td^{n-1} u}\delta_{st}\xbar\sigma^0,
\ee
which can be considered as $2^{n-1}$ supertranslations.

The physical processes that create the impulsive transitions of $\xbar\sigma^0$ and $\Psi_0^i$ are similar to the transitions in \cite{Compere:2019odm}. They are given by shock wave induced by matter fields at different orders.\footnote{One should notice that the matter fields are located outside the region where we are solving the Einstein equation.} The common interpretation of the characteristic initial data problem in NP formalism is that $\p_u \sigma^0$ and $\p_u \xbar\sigma^0$ are the news functions whose time evolutions are not constrained from Einstein equation. Then the time evolutions of all orders of $\Psi_0^i$ are determined once $\xbar\sigma^0$ is given as what we applied previously. However this procedure can also go backwards once $\Psi_0^i$ is given with an impulsive transition $\Psi_0^i + \Theta(u)\delta_{st}\xbar\sigma^0$. For $\Psi_0^j$ that $j>i$, their time evolutions are determined since $\Psi_0^i$ is given. For $\xbar\sigma^0$ and $\Psi_0^k$ that $k<i$, they are completely determined through the time evolution equation \eqref{p0j}. Finally, the transition $\xbar\sigma^0 + \delta\xbar\sigma^0$ will be determined by
\be
(-)^{i-1}\prod_{k=0}^{i-1}\left[\frac{1}{k+1}[\xbar\eth\eth+(k+5)k] \right] \eth^4 \delta \xbar \sigma^0 = \frac{\td \Theta(u)}{\td^{i+2} u}\delta_{st} \xbar\sigma^0 \,.
\ee

\section{Infinite towers of conservation laws}

Memories have their intrinsic connections to conserved quantities. For instance, the leading memory is related to Bondi mass aspect \cite{Strominger:2014pwa} while the subleading memory is related to angular momentum flux \cite{Pasterski:2015tva}. The infinite towers of gravitational memories in \cite{Compere:2019odm} are related to the Noether charges of the infinite towers of supertranslations and superrotations. In the NP formalism, there are infinite conserved quantities\footnote{ In the full gravity theory, there would be flux-balance laws instead of conserved quantities.} derived from the evolution equations of the components of the Weyl tensor at each order \cite{Newman:1968uj}. Equation \eqref{p20} leads to the conservation of mass
\be
\p_u\int\td z \td\bz\,\gamma_{z\bz}\;\Preind_{0}{Y}_{0,0}\,\Psi_2^0=-\p_u\int\td z \td\bz\,\gamma_{z\bz}\;\Preind_{0}{Y}_{0,0}\,\eth^2\xbar\sigma^0=0\,,\label{masslaw}
\ee
where the relations in \eqref{zero} have been used. Equation \eqref{p0j} yields the infinite amount of subleading conservation laws as
\be
\begin{split}
&(n+1)\p_u\int\td z \td\bz\,\gamma_{z\bz}\;\Preind_{2}{Y}_{n-k+2,m}\,\Psi_0^{n+1}\\
=&-\int\td z \td\bz\,\gamma_{z\bz}\;\Preind_{2}{Y}_{n-k+2,m}\,\big(\xbar\eth\eth+(n+5)n\big)\Psi_0^n\\
=&-\int\td z \td\bz\,\gamma_{z\bz}\;\Preind_{2}{Y}_{n-k+2,m}\,k(2n-k+5)\Psi_0^n\,,
\end{split}
\ee
hence
\be
\p_u\int\td z \td\bz\,\gamma_{z\bz}\;\Preind_{2}{Y}_{n+2,m}\,\Psi_0^{n+1}=0\,,\label{newlaws}
\ee
when applying the relations in \eqref{zero}. Equations \eqref{p10} and \eqref{p00} only induce identically vanishing quantities
\begin{align}
\p_u\int\td z \td\bz\,\gamma_{z\bz}\;\Preind_{0}{Y}_{0,0}\,\xbar\eth\Psi_1^0&=-\int\td z \td\bz\,\gamma_{z\bz}\;\Preind_{0}{Y}_{0,0}\,\xbar\eth\eth\Psi_2^0=0\,,\label{trivial1}\\
\p_u\int\td z \td\bz\,\gamma_{z\bz}\;\Preind_{0}{Y}_{0,0}\,\xbar\eth^2\Psi_0^0&=-\int\td z \td\bz\,\gamma_{z\bz}\;\Preind_{0}{Y}_{0,0}\,\xbar\eth^2\eth\Psi_1^0=0\,.\label{trivial2}
\end{align}

The conserved quantities can be reorganized as a unique charge in expansion (see, for instance, \cite{Compere:2017wrj,Godazgar:2018vmm,Godazgar:2018qpq,Godazgar:2018dvh} for relevant developments). The supertranslation charge at each order proposed in \cite{Conde:2016rom} is given by
\footnote{A factor $\frac12$ is missing on the righthand side of both eq.(2.26) and eq.(2.27) in \cite{Conde:2016rom}. To compare the results in \cite{Conde:2016rom} to those in \cite{Newman:1968uj}, one should notice that, in \cite{Conde:2016rom}, the signature is $(-,+,+,+)$ and a unit sphere boundary is chosen. Moreover one needs to do the rescaling $\Psi_0^{(m)}\rightarrow\frac{2}{(m+3)(m+2)}\Psi_0^{(m)}\,(m\geq1)$.}
\begin{multline}\label{charge}
Q_{st}=\int \td z \td\bz\,\ga\;T(z,\bz)\,\bigg[\Psi_2^0+\xbar\Psi_2^0-\frac{\eth\xbar\Psi_1^0+\xbar\eth\Psi_1^0}{3r} + \frac{\eth^2\xbar\Psi_0^0+\xbar\eth^2\Psi_0^0}{12r^2}\\
+\sum\limits_{m=1}^\infty\frac{1}{(m+1)(m+4)}\frac{\eth^2\xbar\Psi_0^{(m)}+\xbar\eth^2{\Psi}_0^{(m)}}{r^{m+2}}\bigg]\,.
\end{multline}
All the conserved quantities in equations \eqref{masslaw}-\eqref{trivial2} can be recovered from certain order of the supertranslation charge \eqref{charge}. It is worth to point out that the charge \eqref{charge} was proposed from the connections to soft graviton theorems. It is not derived from action principle of the theory. Recently, BMS charges at different orders were studied carefully by covariant phase space methods \cite{Godazgar:2018vmm}. At the leading order, the charge in \cite{Godazgar:2018vmm} agrees with \eqref{charge}, while the subleading order charge in \cite{Godazgar:2018vmm} is absent for vacuum gravity. At the third order, the charge in \cite{Godazgar:2018vmm} is much more complicated than that in \eqref{charge}. Although the subleading charges in \cite{Godazgar:2018vmm} are different from \eqref{charge}, they do have intrinsic connections to the Newman-Penrose conserved quantities reviewed previously in this section. Since the charge \eqref{charge} can be understood as a recast from soft theorems in the low energy expansion to a charge in the $\frac1r$ expansion. It is possible that the NP conserved quantities, i.e. the specially selected supertranslation charge, are the essence to recover the soft theorems rather than the full supertranslation charge. Nonetheless, defining conserved charges to asymptotic symmetries is a well-established issue \cite{Wald:1999wa,Barnich:2001jy}, extending those charges to the subleading orders is still a tricky question.

\section{Comment on the relation to soft theorems}

Memories and soft theorems are mathematically equivalent in the context of the triangle relation \cite{Strominger:2017zoo}. However such relation is not easy to verify beyond the subsubleading order since the soft graviton theorems beyond the subsubleading order do not have an universal factorization property \cite{Cachazo:2014fwa,Hamada:2018vrw}. Correspondingly, the higher order memory formulas \eqref{subs} are in more complicated forms. The first two orders of memories are shown to be equivalent to the leading soft graviton theorems \cite{Strominger:2014pwa} and subleading soft graviton theorem \cite{Pasterski:2015tva}. The key observation of the equivalence is the fact that the soft factors, after Fourier transform and projection on the null infinity, can be considered as classical fields that satisfy classical equations of motion, namely the Einstein equation. Hence the equivalence between the first two orders of memories and soft theorems can be interpreted as \cite{Strominger:2014pwa,Pasterski:2015tva}
\be
S^{(0)}_{zz} \sim \ga \xbar \sigma^0 \,,\quad S^{(1)}_{zz} \sim \int \td u' \, \ga \xbar \sigma^0 \, .
\ee
Including local stress-energy tensor, the classical equations of motion\footnote{Here we follow the results in \cite{Conde:2016rom} in the signature $(-,+,+,+)$ in order to compare to the computation in \cite{Strominger:2014pwa,Pasterski:2015tva}.} that the leading factor and the subleading soft factor satisfy are
\be
\Psi_2^0 + \xbar \Psi_2^0 = \eth^2\p_u \xbar\sigma^0 + \xbar \eth^2\p_u \sigma^0 +\frac{1}{2}\int \td u'\, T^0_{uu}\,,
\ee
which is the real (electric) part of
\be
\p_u \Psi_2^0=\eth^2\p_u \xbar\sigma^0 + \frac{1}{4} T^0_{uu}\,,
\ee
and
\be
\xbar\eth \Psi_1^0 - \eth \xbar \Psi_1^0 = \xbar\eth\eth\int\td u' \, (\eth^2\xbar\sigma^0 - \xbar\eth^2 \sigma^0) + \int \td u'\,\left(\xbar\eth \frac{T^0_{u\bz}}{2\sga} - \eth \frac{T^0_{uz}}{2\sga}\right)\,,
\ee
which is the imaginary (magnetic) part of\footnote{The relation $\Psi_2^0 - \xbar \Psi_2^0 = \eth^2\xbar\sigma^0 - \xbar\eth^2 \sigma^0$ needs to be utilized.}
\be
\p_u \Psi_1^0=\eth\Psi_2^0 + \frac{1}{2\sga} T^0_{u\bz}\,.\label{Phi1}
\ee

Regarding to the third memory, it is supposed to be equivalent to the subsubleading soft factor discovered in \cite{Cachazo:2014fwa}. After Fourier transform and projection on the null infinity, the subsubleading soft factor should be interpreted as classical field
\be
S^{(2)}_{zz} \sim \int \td u' \td u''\, \ga \xbar \sigma^0 \, ,
\ee
and the classical equation of motion for the subsubleading soft factor should be related to
\be
\p_u\Psi_0^0=\eth\Psi_1^0+\frac{3}{4\ga}T_{\bz\bz}^0\,.\label{Phi0}
\ee
Equation \eqref{Phi0} can be re-organized as
\begin{multline}\label{S2}
\xbar\eth^2 \Psi_0^0 - \eth^2 \xbar \Psi_0^0 = \xbar\eth^2\eth^2\int\td u' \td u''\, (\eth^2\xbar\sigma^0 - \xbar\eth^2 \sigma^0)\\ + \int \td u' \td u''\,\left(\xbar\eth^2 \eth \frac{T^0_{u\bz}}{2\sga} - \eth^2\xbar\eth \frac{T^0_{uz}}{2\sga}\right)\\  + \int \td u'\,\left(\xbar\eth^2 \frac{3T^0_{\bz\bz}}{4\sga} - \eth^2 \frac{3T^0_{zz}}{4\sga}\right)\,,
\end{multline}
to connect to the expression $\int \td u' \td u''\, \ga \xbar \sigma^0 \,$. However, considering stress-energy tensor from massless particles or localized wave packets which puncture the null infinity at points $(u_k,z_k,\bz_k)$ as in \cite{Pasterski:2015tva}, direct computation shows that equation \eqref{S2} does not fulfill when $\int \td u' \td u''\, \ga \xbar \sigma^0 \,$ is replaced by the subsubleading soft factor. The connection at the third order seems to be still of a mystery. However the failure of demonstrating the equivalence between gravitational memories and soft graviton theorems beyond subleading order may not be surprising in the sense that the soft graviton theorems beyond subleading order may have different natures. On the one hand, the soft graviton theorems beyond subsubleading order do not have a universal property \cite{Hamada:2018vrw} in contrast to the first three orders. On the other hand, the double copy relation \cite{Kawai:1985xq} indicates that a gravity amplitude can be expressed as a sum of the square of color-ordered Yang-Mills amplitudes. The first three soft factors in gravity satisfy precisely the double copy relation \cite{He:2014bga,Liu:2014vva}. From this point of view, the third order gravitational memory should have intrinsic relation to the first two orders. Unfortunately, we are not able to uncover that in the present work. It is definitely an interesting question that should be addressed elsewhere.

\section{Conclusion}

We demonstrated that there are infinite towers of gravitational memories which can be associated to the same supertranslation in the NP formalism. An infinite number of conserved quantities can be obtained from each order of the supertranslation charge. We also commented on the relation between the infinite towers of gravitational memories and the infinite towers of soft graviton theorems. The equivalence of those two subjects is not yet clear beyond the second order.

As a final remark, it is worthwhile to point out that the analysis performed in gravitational theory can be easily extended to electromagnetism by applying the results in \cite{Newman:1968uj} and \cite{Conde:2016csj}.

\section*{Acknowledgments}

The author thanks Geoffrey Comp\`{e}re for useful discussions. This work is supported in part by the National Natural Science Foundation of China under Grants No. 11905156 and No. 11935009.

\appendix

\section{Spin-weighted derivative operators}
\label{eth}

The operators $\eth$ and $\xbar\eth$ were originally introduced in~\cite{Newman:1966ub} to replace the covariant derivative on a sphere with metric $\td S^2 =2 \ga\, \td z\td \bz\,$. The choice in \cite{Newman:1968uj} is $\ga=\frac{1}{(1+z\bz)^2}$ which is not the unit sphere. The definitions of $\eth$ and $\xbar\eth$ on a field $\eta$ are
\be
\eth \eta=\ga^{\frac{s-1}{2}}\pzb (\eta\ga^{-\frac{s}{2}})\,,\;\;\;\; \xbar\eth \eta=\ga^{\frac{-s-1}{2}}\pz (\eta\ga^{\frac{s}{2}})\,,
\ee
where $s$ is the spin weight of $\eta$. The spin weight of the relevant fields are listed in Table \ref{t1}.
\begin{table}[h]
\caption{Spin weights}\label{t1}
\begin{center}
\begin{tabular}{c|c|c|c|c|c|c|c|c} & $\sigma^0$ & $\xbar\sigma^0$ & $\p_u \xbar\sigma^0$ &$\Psi^0_2$ & $\Psi^0_1$ &$\Psi_0^0$ & $\Psi_0^{(m)}$  &  \\
\hline
s &  $2$ &  $-2$ &  $-2$  & $0$ & $1$ & $2$  & $2$ & \\
\end{tabular}
\end{center} \end{table}

The operator $\eth$ ($\xbar\eth$) will increase (decrease) the spin weight. These two operators do not commute in general. Their commutation relation is
\be
[\xbar\eth,\eth]\eta=\frac{s}{2}R_S\eta\,,
\ee
where $R_S$ is the Ricci scalar of the sphere.

Spin $s$ spherical harmonics are defined by acting $\eth$ and $\xbar\eth$ on the spherical harmonics $Y_{l,m}$ $(l=0,1,2,\ldots\,;\,m=-l,\ldots,l)$. They are given by
\begin{equation}
	\Preind_{s}{Y}_{l,m}=\begin{cases}
	\sqrt{\dfrac{(l-s)!}{(l+s)!}}\,\eth^s Y_{l,m}\qquad\qquad(0\leq s\leq l)\\
	(-1)^s\sqrt{\dfrac{(l+s)!}{(l-s)!}}\,\xbar\eth^{-s} Y_{l,m}\quad(-l\leq s\leq 0)
\end{cases} \,.
\end{equation}
From the definition of spin weighted spherical harmonics, one can obtain that
\begin{equation}
\label{zero}
\begin{gathered}
	\int\td z \td\bz\,\gamma_{z\bz}\, \Preind_{s}{Y}_{l,m}\,\eth^{l-s+1}\eta =0 \,,\\
	\int \td z \td\bz\,\gamma_{z\bz}\,\Preind_{s}{\bar{Y}}_{l,m}  \,\xbar\eth^{l-s+1}\zeta=0 \,,\\
	\xbar\eth\eth\Preind_{s}{Y}_{l,m}= - (l-s)(l+s+1)\Preind_{s}{Y}_{l,m} \,, \\
	\int\td z \td\bz\,\gamma_{z\bz}\, A\eth B=-\int\td z \td\bz\,\gamma_{z\bz}\, B\eth A \,.
\end{gathered}
\end{equation}
where $\eta$ and $\zeta$ are with spin weight $-l-1$ and $l+1$ respectively, and the expression $A\eth B$ has spin weight zero.

\bibliography{ref}

\providecommand{\href}[2]{#2}\begingroup\raggedright\begin{thebibliography}{10}

\bibitem{Compere:2019odm}
G.~Comp\`{e}re, ``{Infinite towers of supertranslation and superrotation
  memories},'' \href{http://dx.doi.org/10.1103/PhysRevLett.123.021101}{{\em
  Phys. Rev. Lett.} {\bfseries 123} no.~2, (2019) 021101},
\href{http://arxiv.org/abs/1904.00280}{{\ttfamily arXiv:1904.00280 [gr-qc]}}.

\bibitem{memory}
Y.~B. Zel'dovich and A.~G. Polnarev, ``{Radiation of gravitational waves by a
  cluster of superdense stars},'' {\em Soviet Astronomy} {\bfseries 18} (Aug.,
  1974) 17.

\bibitem{Braginsky:1986ia}
V.~B. Braginsky and L.~P. Grishchuk, ``{Kinematic Resonance and Memory Effect
  in Free Mass Gravitational Antennas},'' {\em Sov. Phys. JETP} {\bfseries 62}
  (1985) 427--430.
[Zh. Eksp. Teor. Fiz.89,744(1985)].

\bibitem{1987Natur}
V.~B. {Braginskii} and K.~S. {Thorne}, ``{Gravitational-wave bursts with memory
  and experimental prospects},'' \href{http://dx.doi.org/10.1038/327123a0}{{\em
  Nature} {\bfseries 327} (May, 1987) 123--125}.

\bibitem{Christodoulou:1991cr}
D.~Christodoulou, ``{Nonlinear nature of gravitation and gravitational wave
  experiments},''
\href{http://dx.doi.org/10.1103/PhysRevLett.67.1486}{{\em Phys. Rev. Lett.}
  {\bfseries 67} (1991) 1486--1489}.

\bibitem{Strominger:2014pwa}
A.~Strominger and A.~Zhiboedov, ``{Gravitational Memory, BMS Supertranslations
  and Soft Theorems},'' \href{http://dx.doi.org/10.1007/JHEP01(2016)086}{{\em
  JHEP} {\bfseries 01} (2016) 086},
  \href{http://arxiv.org/abs/1411.5745}{{\ttfamily arXiv:1411.5745 [hep-th]}}.

\bibitem{Strominger:2017zoo}
A.~Strominger, ``{Lectures on the Infrared Structure of Gravity and Gauge
  Theory},''
\href{http://arxiv.org/abs/1703.05448}{{\ttfamily arXiv:1703.05448 [hep-th]}}.

\bibitem{Hamada:2018vrw}
Y.~Hamada and G.~Shiu, ``{Infinite Set of Soft Theorems in Gauge-Gravity
  Theories as Ward-Takahashi Identities},''
  \href{http://dx.doi.org/10.1103/PhysRevLett.120.201601}{{\em Phys. Rev.
  Lett.} {\bfseries 120} no.~20, (2018) 201601},
  \href{http://arxiv.org/abs/1801.05528}{{\ttfamily arXiv:1801.05528
  [hep-th]}}.

\bibitem{Newman:1961qr}
E.~Newman and R.~Penrose, ``{An Approach to gravitational radiation by a method
  of spin coefficients},''
\href{http://dx.doi.org/10.1063/1.1724257}{{\em J. Math. Phys.} {\bfseries 3}
  (1962) 566--578}.

\bibitem{Newman:1968uj}
E.~Newman and R.~Penrose, ``{New conservation laws for zero rest-mass fields in
  asymptotically flat space-time},''
  \href{http://dx.doi.org/10.1098/rspa.1968.0112}{{\em Proc. Roy. Soc. Lond. A}
  {\bfseries A305} (1968) 175--204}.

\bibitem{Conde:2016rom}
E.~Conde and P.~Mao, ``{BMS Supertranslations and Not So Soft Gravitons},''
  \href{http://dx.doi.org/10.1007/JHEP05(2017)060}{{\em JHEP} {\bfseries 05}
  (2017) 060},
\href{http://arxiv.org/abs/1612.08294}{{\ttfamily arXiv:1612.08294 [hep-th]}}.

\bibitem{Pasterski:2015tva}
S.~Pasterski, A.~Strominger, and A.~Zhiboedov, ``{New Gravitational
  Memories},'' \href{http://dx.doi.org/10.1007/JHEP12(2016)053}{{\em JHEP}
  {\bfseries 12} (2016) 053}, \href{http://arxiv.org/abs/1502.06120}{{\ttfamily
  arXiv:1502.06120 [hep-th]}}.

\bibitem{Thorne:1992sdb}
K.~S. Thorne, ``{Gravitational-wave bursts with memory: The Christodoulou
  effect},''
\href{http://dx.doi.org/10.1103/PhysRevD.45.520}{{\em Phys. Rev.} {\bfseries
  D45} no.~2, (1992) 520--524}.

\bibitem{Flanagan:2005yc}
E.~E. Flanagan and S.~A. Hughes, ``{The Basics of gravitational wave theory},''
  \href{http://dx.doi.org/10.1088/1367-2630/7/1/204}{{\em New J. Phys.}
  {\bfseries 7} (2005) 204},
  \href{http://arxiv.org/abs/gr-qc/0501041}{{\ttfamily arXiv:gr-qc/0501041}}.

\bibitem{ValienteKroon:1998vn}
J.~A. Valiente~Kroon, ``{Logarithmic Newman-Penrose constants for arbitrary
  polyhomogeneous space-times},''
  \href{http://dx.doi.org/10.1088/0264-9381/16/5/314}{{\em Class. Quant. Grav.}
  {\bfseries 16} (1999) 1653--1665},
  \href{http://arxiv.org/abs/gr-qc/9812004}{{\ttfamily arXiv:gr-qc/9812004}}.

\bibitem{Barnich:2011ty}
G.~Barnich and P.-H. Lambert, ``{A Note on the Newman-Unti group and the BMS
  charge algebra in terms of Newman-Penrose coefficients},''
  \href{http://dx.doi.org/10.1155/2012/197385}{{\em J. Phys. Conf. Ser.}
  {\bfseries 410} (2013) 012142},
  \href{http://arxiv.org/abs/1102.0589}{{\ttfamily arXiv:1102.0589 [gr-qc]}}.

\bibitem{Newman:1962cia}
E.~T. Newman and T.~W.~J. Unti, ``{Behavior of Asymptotically Flat Empty
  Spaces},''
\href{http://dx.doi.org/10.1063/1.1724303}{{\em J. Math. Phys.} {\bfseries 3}
  no.~5, (1962) 891}.

\bibitem{Compere:2017wrj}
G.~Comp\`{e}re, R.~Oliveri, and A.~Seraj, ``{Gravitational multipole moments
  from Noether charges},''
  \href{http://dx.doi.org/10.1007/JHEP05(2018)054}{{\em JHEP} {\bfseries 05}
  (2018) 054}, \href{http://arxiv.org/abs/1711.08806}{{\ttfamily
  arXiv:1711.08806 [hep-th]}}.

\bibitem{Godazgar:2018vmm}
H.~Godazgar, M.~Godazgar, and C.~Pope, ``{Subleading BMS charges and fake news
  near null infinity},'' \href{http://dx.doi.org/10.1007/JHEP01(2019)143}{{\em
  JHEP} {\bfseries 01} (2019) 143},
  \href{http://arxiv.org/abs/1809.09076}{{\ttfamily arXiv:1809.09076
  [hep-th]}}.

\bibitem{Godazgar:2018qpq}
H.~Godazgar, M.~Godazgar, and C.~Pope, ``{New dual gravitational charges},''
  \href{http://dx.doi.org/10.1103/PhysRevD.99.024013}{{\em Phys. Rev. D}
  {\bfseries 99} no.~2, (2019) 024013},
  \href{http://arxiv.org/abs/1812.01641}{{\ttfamily arXiv:1812.01641
  [hep-th]}}.

\bibitem{Godazgar:2018dvh}
H.~Godazgar, M.~Godazgar, and C.~Pope, ``{Tower of subleading dual BMS
  charges},'' \href{http://dx.doi.org/10.1007/JHEP03(2019)057}{{\em JHEP}
  {\bfseries 03} (2019) 057}, \href{http://arxiv.org/abs/1812.06935}{{\ttfamily
  arXiv:1812.06935 [hep-th]}}.

\bibitem{Wald:1999wa}
R.~M. Wald and A.~Zoupas, ``{A General definition of 'conserved quantities' in
  general relativity and other theories of gravity},''
  \href{http://dx.doi.org/10.1103/PhysRevD.61.084027}{{\em Phys. Rev. D}
  {\bfseries 61} (2000) 084027},
  \href{http://arxiv.org/abs/gr-qc/9911095}{{\ttfamily arXiv:gr-qc/9911095}}.

\bibitem{Barnich:2001jy}
G.~Barnich and F.~Brandt, ``{Covariant theory of asymptotic symmetries,
  conservation laws and central charges},''
  \href{http://dx.doi.org/10.1016/S0550-3213(02)00251-1}{{\em Nucl. Phys. B}
  {\bfseries 633} (2002) 3--82},
  \href{http://arxiv.org/abs/hep-th/0111246}{{\ttfamily arXiv:hep-th/0111246}}.

\bibitem{Cachazo:2014fwa}
F.~Cachazo and A.~Strominger, ``{Evidence for a New Soft Graviton Theorem},''
  \href{http://arxiv.org/abs/1404.4091}{{\ttfamily arXiv:1404.4091 [hep-th]}}.

\bibitem{Kawai:1985xq}
H.~Kawai, D.~Lewellen, and S.~Tye, ``{A Relation Between Tree Amplitudes of
  Closed and Open Strings},''
  \href{http://dx.doi.org/10.1016/0550-3213(86)90362-7}{{\em Nucl. Phys. B}
  {\bfseries 269} (1986) 1--23}.

\bibitem{He:2014bga}
S.~He, Y.-t. Huang, and C.~Wen, ``{Loop Corrections to Soft Theorems in Gauge
  Theories and Gravity},''
  \href{http://dx.doi.org/10.1007/JHEP12(2014)115}{{\em JHEP} {\bfseries 12}
  (2014) 115}, \href{http://arxiv.org/abs/1405.1410}{{\ttfamily arXiv:1405.1410
  [hep-th]}}.

\bibitem{Liu:2014vva}
Z.-W. Liu, ``{Soft theorems in maximally supersymmetric theories},''
  \href{http://dx.doi.org/10.1140/epjc/s10052-015-3304-1}{{\em Eur. Phys. J. C}
  {\bfseries 75} no.~3, (2015) 105},
  \href{http://arxiv.org/abs/1410.1616}{{\ttfamily arXiv:1410.1616 [hep-th]}}.

\bibitem{Conde:2016csj}
E.~Conde and P.~Mao, ``{Remarks on asymptotic symmetries and the subleading
  soft photon theorem},''
  \href{http://dx.doi.org/10.1103/PhysRevD.95.021701}{{\em Phys. Rev. D}
  {\bfseries 95} no.~2, (2017) 021701(R)},
  \href{http://arxiv.org/abs/1605.09731}{{\ttfamily arXiv:1605.09731
  [hep-th]}}.

\bibitem{Newman:1966ub}
E.~T. Newman and R.~Penrose, ``{Note on the Bondi-Metzner-Sachs group},''
\href{http://dx.doi.org/10.1063/1.1931221}{{\em J. Math. Phys.} {\bfseries 7}
  (1966) 863--870}.

\end{thebibliography}\endgroup

\end{document}